\documentclass[journal,twoside,web]{ieeecolor}
\usepackage{generic}
\usepackage{cite}
\usepackage{amsmath,amssymb,amsfonts}
\usepackage{algorithm}  
\usepackage{algorithmicx}  
\usepackage{algpseudocode}  
\usepackage{graphicx}
\usepackage{amsmath}
\usepackage{autobreak}
\usepackage{subfigure}
\def\BibTeX{{\rm B\kern-.05em{\sc i\kern-.025em b}\kern-.08em
    T\kern-.1667em\lower.7ex\hbox{E}\kern-.125emX}}
\begin{document}
\title{G-flocking: Flocking Model Optimization based on Genetic Framework}
\author{Li Ma, Weidong Bao, Xiaomin Zhu, \IEEEmembership{Member, IEEE}, Meng Wu, Yuan Wang, Yunxiang Ling, and Wen Zhou
\thanks{This work was supported in part by the National Natural Science Foundation of China under Grants 61872378,  91648204, 71702186, in part by Postgraduate Research Innovation Project in Hunan Province under grant CX2018B021, in part by the Scientific Research Project of National University of Defense Technology under Grants  ZK17-03-48}}
\maketitle

\begin{abstract}
Flocking model has been widely used to control robotic swarm. However, with the increasing scalability, there exist complex conflicts for robotic swarm in autonomous navigation, brought by internal pattern maintenance, external environment changes, and target area orientation, which results in poor stability and adaptability. Hence, optimizing the flocking model for robotic swarm in autonomous navigation is an important and meaningful research domain.
\end{abstract}


\section{Introduction}
\label{sec:introduction}
\IEEEPARstart{R}{obotic} swarms pose an attractive and scalable solution to accomplish complicated missions such as search-and-rescue\cite{b1}, mapping\cite{b2}, target tracking\cite{b3}, and full coverage attacking, which can prevent human beings from boring, harsh and dangerous environment. One main advantage of a robotic swarm solution is the simple local interactions between individuals within a complex system that could generate some new properties and phenomena observed at the system level, such as a set of collective behaviors\cite{b4}. Much like their biological counterparts such as fish schools\cite{b5}, bird flocks\cite{b6}, ant colonies\cite{b7}, and cell populations\cite{b8}, the resulting collective patterns are robust and flexible to agents joining in and dropping out, especially when accidents like obstacles, dangers, and new missions emerge. While robotic swarm enjoys numerous advantages, the large-scale autonomous robotic swarm incurs a high robot-failure probability due to real-life conditions when delays, uncetainties, and kinematic constraints are present. This phenomenon is even more noticeable in military projects like Gremlins\cite{b9} and LOCUST\cite{b10}, since they are built on small, low-cost and semi-autonomous UAVs whose failure probability is expected to be much higher.

Among researches of flocking models with obstacle avoidance, Wang et al.\cite{b11} proposed an improved fast flocking algorithm with obstacle avoidance for multi-agent dynamic systems based on Olfati-Sabers algorithm. Li et al.\cite{b12} studied the flocking problem of multi-agent systems with obstacle avoidance, in the situation when only a fraction of the agents have information on the obstacles. Vrohidis et al.\cite{b13} considered a networked multi-robot system operating in an obstacle populated planar workspace under a single leader-multiple followers architecture.

\begin{figure}[H]
	\centerline{\includegraphics[width=0.9\columnwidth]{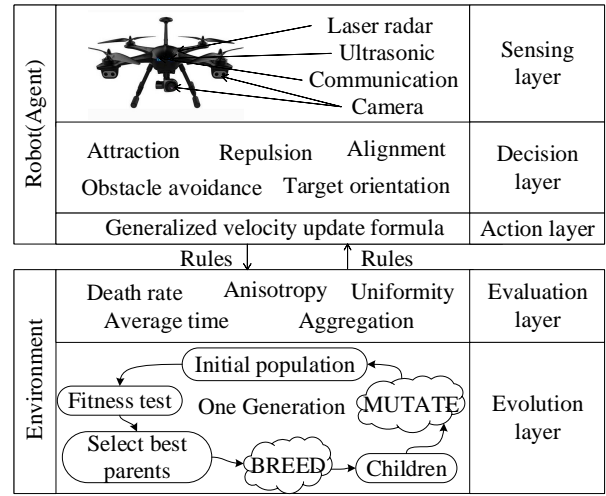}}
	\caption{Genetic flocking optimizing framework}
	\label{fig1}
\end{figure}

To the best of our knowledge, few previous literatures studied the model that satisfies both stability and adaptivity of the autonomous robotic swarm. Thus, we design a novel genetic flocking optimizing framework that can achieve both stability and adaptivity of the robotic swarms. As shown in Fig. 1, a robot is seperated into three layers, including sensing layer, decision layer, and action layer, which supports the basic navigation function. Besides, we generalize velocity updating formula by rules described with weight parameters, which evolves through interaction with the environment. The environment is divided into two layers: the evaluation layer and the evolutionary layer, where the former provides fitness function for the latter. 

\section{GENERALIZED FLOCKING MODEL}

As shown in Fig. 2, a robot agent $i$ has four detection areas: repulsion area, alignment area, attraction area, and obstacle avoidance area. The velocity updating fomula is described as follows:

In equation (1), we define the weight parameters ${\rm{a, b, c, d, e}} \in \left( {0,1} \right)$, which is used to flexibly handle the generalization formula. 

\begin{figure}[H]
\centerline{\includegraphics[width=0.7\columnwidth]{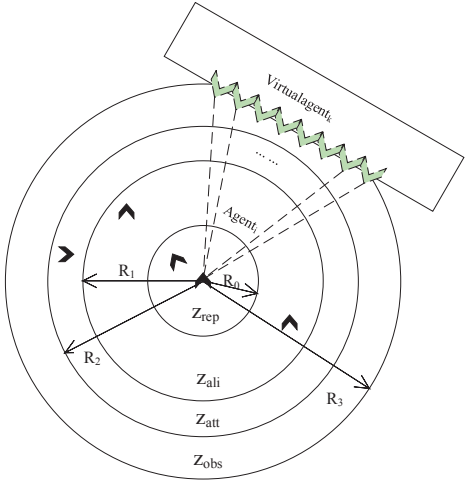}}
\caption{Pattern-formation areas ($Z_{rep}, Z_{ali}, Z_{att}$) and obstacle-avoidance area ($Z_{obs}$). }\end{figure}

\begin{equation}
\begin{split}
\Delta v_i &= 
a\sum\limits_{j \in {Z_{rep}}} {({R_0} - {r_{ij}})} \frac{{{p_i} - {p_j}}}{{{r_{ij}}}} 
\\&+ b\frac{1}{{{N_{ali}}}}\sum\limits_{j \in {Z_{ali}}} {\frac{{{v_j}}}{{\left| {{v_j}} \right|}} + c} \sum\limits_{j \in {Z_{att}}} {({R_2} - {r_{ij}})} \frac{{{p_j} - {p_i}}}{{{r_{ij}}}} 
\\&+ d\sum\limits_{j \in {Z_{obs}}} {({R_3} - {r_{ik}})\frac{{{p_i} - {p_k}}}{{{r_{ik}}}} + e\frac{{{p_{tar}} - {p_i}}}{{{r_{itar}}}} + v(t)} .
\end{split}
\end{equation}

Tunning the model above means that we propose four rules referring to classic reynolds' boids model and we optimize the parameters there. Note that the parameter space is 20-dimensional; therefore, manual tuning, global optimization methods, or parameter sweeping would be generally much time-consuming.

\section{Order Parameters and Fitness Function}
In order to select the set of parameters that perform the best in the simulation process, we propose a fitness function composed by several order parameters, which helps us to abstract the mathematical model of single objective optimization. The fitness function is described as flollows:

\begin{equation}\begin{split}F &= \frac{{\sum\nolimits_j {(T_j^{arrive} - T_j^{start})} }}{N} \times \frac{{{N_{death}}}}{{{N_{total}}}} \\&\times \frac{{\sum\nolimits_t {\sum\nolimits_j {\sqrt {{{(p_j^x - r_t^x)}^2} + {{(p_j^y - r_t^y)}^2}} } } }}{{NT}} \\&\times \frac{{\sum\nolimits_t {({\gamma _t} - \bar \gamma )} }}{T} \\&\times \frac{{\sum\nolimits_t {\sum\nolimits_j {{{(\theta _j^t - {\delta ^t})}^2}} } }}{T} \times \alpha  ,\end{split}\end{equation}

where $T_j^{start}$ is the time when the navigation is triggered, and $T_j^{arrive}$ is the time when robotic agent $j$ reaches the target area. $N_{death}$ represents the number of the dead agent, and $N_{total}$ represents the number of the total agents in the robotic swarm. $r_t^{x}$ and $r_t^{y}$ are the abscissa and ordinate of the position of the swarm's centroid at time $t$. $T$ is the total time of the whole navigation process, while $N$ is the agent number of the agents that arriving at target area.

We define the stability of the robotic swarm as the variance of the $\gamma_{t}$ sequence, which describes whether the flock structure of this swarm is stable.
\begin{equation}s_\gamma ^2 = \frac{{\sum\nolimits_{t = 0}^T {{{\left( {{\gamma _t} - \bar \gamma } \right)}^2}} }}{T},\end{equation}
\begin{equation}{\gamma _t} = \frac{{\sum\limits_j {\sqrt {{{\left( {p_j^x - r_t^x} \right)}^2} + {{\left( {p_j^y - r_t^y} \right)}^2}} } }}{N}.\end{equation}

We define anisotropic index to describe the variation of population velocity direction. Specifically, it needs to calculate the average angle of each individual velocity direction and flock velocity direction at a certain time, and then calculate the average value of the whole process, which is the index of anisotropic index. The variance of the average angle of the whole process represents the variation range of anisotropic index, and the formula of anisotropy is as follows:
\begin{equation}{\delta ^t} = \frac{{\sum\limits_j {{\theta _j}} }}{N}.\end{equation}

With this method, we created a single-objective optimization scenario, which can be solved using genetic algorithm.

\section{The G-flocking algorithm}
This research adopts Parameter Tuning of Flocking Model based on classical genetic algorithm (GA) framework. The main algorithm is described as follows:

\begin{algorithm}[H]  
	\caption{The G-flocking algorithm}
	\label{alg::conjugateGradient}  
	\begin{algorithmic}[1] 
		\Require 
		$R^{exp}$: a set of traditional expert's rules;
		$P(0)$: randomly generate an initial population of rules;
		$M$: maximum number of iterations;
		$N_{p}$: number of the population;
		$L_{R}$: length of the rules;
		$N_{s}$: number of seed used to produce the next generation;
		$r$: member mutation rate;   
		\Ensure
		$R^{opt}$: an set of optimal rules;    
		\State $t=0$
		\While{$t<=T$}
		\For{$i = 1 \to M $}
		\State Evaluate fitness of $P(t)$
		\EndFor
		\For{$i = 1 \to M $}
		\State Select operation to $P(t)$ 
		\EndFor
		\For{$i = 1 \to M/2 $}
		\State Crossover operation to $P(t) = Crossover(N_{s}, N_{p})$
		\EndFor
		\For{$i = 1 \to M $}
		\State Mutation operation to $P(t) = Mutation(P(t), r)$
		\EndFor
		\For{$i = 1 \to M $}
		\State $P(t+1)=P(t)$
		\EndFor
		\State $t = t + 1$
		\EndWhile
	\end{algorithmic}  
\end{algorithm}

In the algorithm, the random rules $R^{exp}$ are represented as: $\left\{R^{1}_{0},R^{2}_{0},R^{3}_{0},R^{4}_{0}\right\}$, and $R^{i}_{0}=\left\{a^{i}_{0},b^{i}_{0},c^{i}_{0},d^{i}_{0},e^{i}_{0}\right\}, i=1,2,3,4$.

The outputs of G-flocking are also a set of optimized rules:
$R^{opt}=\left\{R^{1}_{opt},R^{2}_{opt},R^{3}_{opt},R^{4}_{opt}\right\}$, and $R^{i}_{opt}=\left\{a^{i}_{opt},b^{i}_{opt},c^{i}_{opt},d^{i}_{opt},e^{i}_{opt}\right\}, i=1,2,3,4$.

Once optimized, we can get the optimal rules for the flocking which composing the BRIAN model.

\section{Experiment Analysis}
To reveal the performance improvements of BRIAN, we compare it with \underline{b}asic \underline{r}ule-bas\underline{e}d \underline{m}odel (BREAM). BREAM derives from the classical Reynolds' flocking model that has been widely used. To apply Reynolds' flocking model to more complex environment, the comprehensive obstacle avoidance strategies are integrated into BREAM. 

\begin{figure}[H]
	\centering
	\subfigure[BREAM (20 robots)]{
		\includegraphics[width=3.6cm]{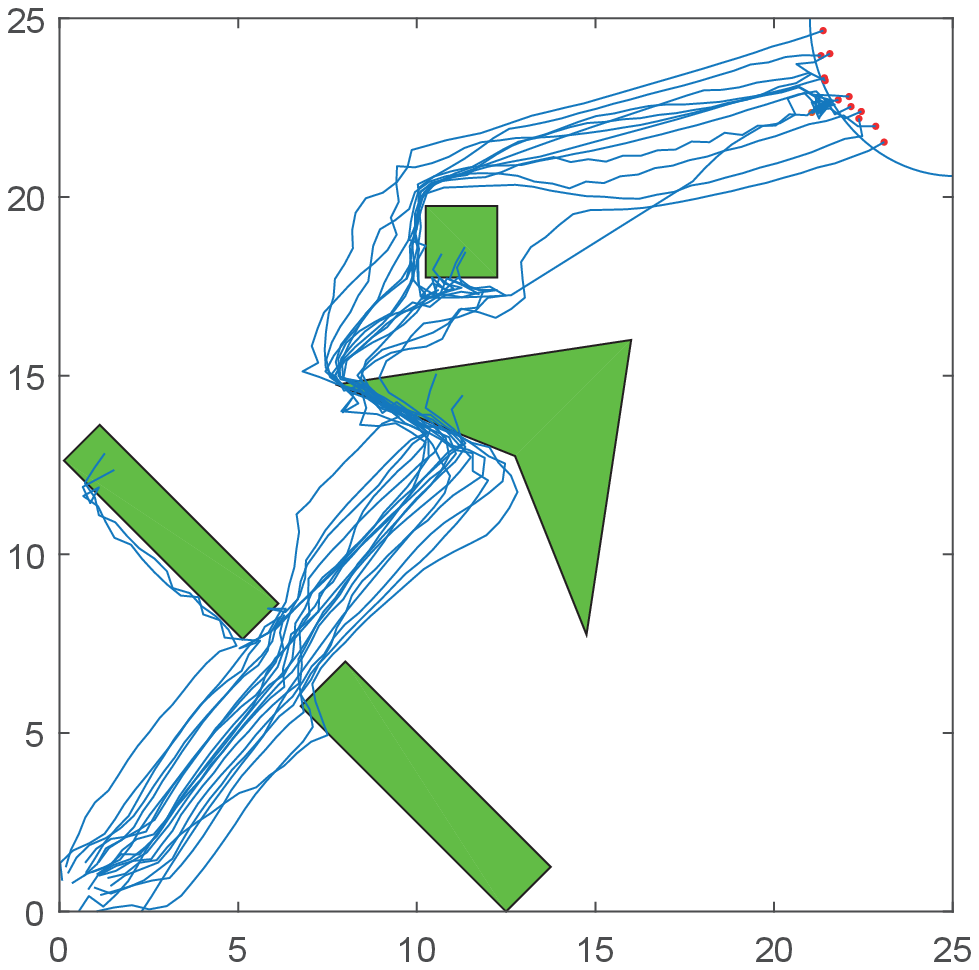}
	}
	\quad
	\subfigure[BRIAN (20 robots)]{
		\includegraphics[width=3.6cm]{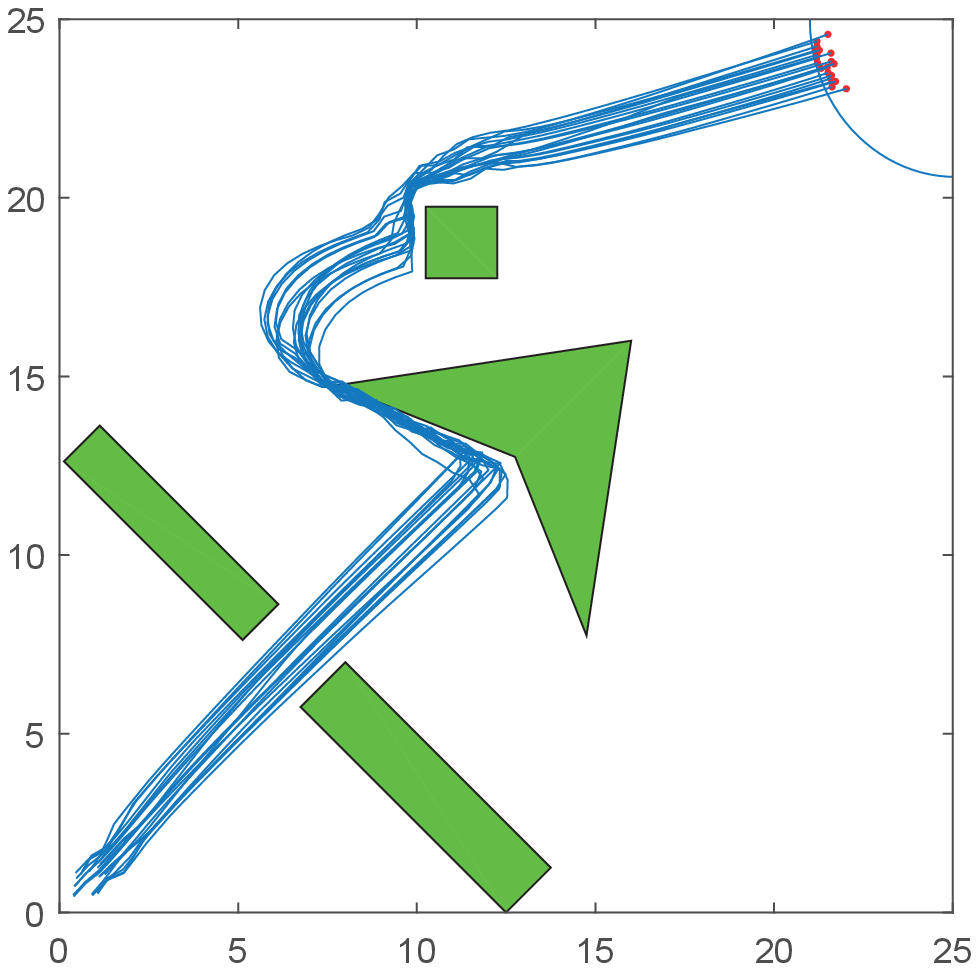}
	}
	\quad
	\subfigure[BREAM (60 robots)]{
		\includegraphics[width=3.6cm]{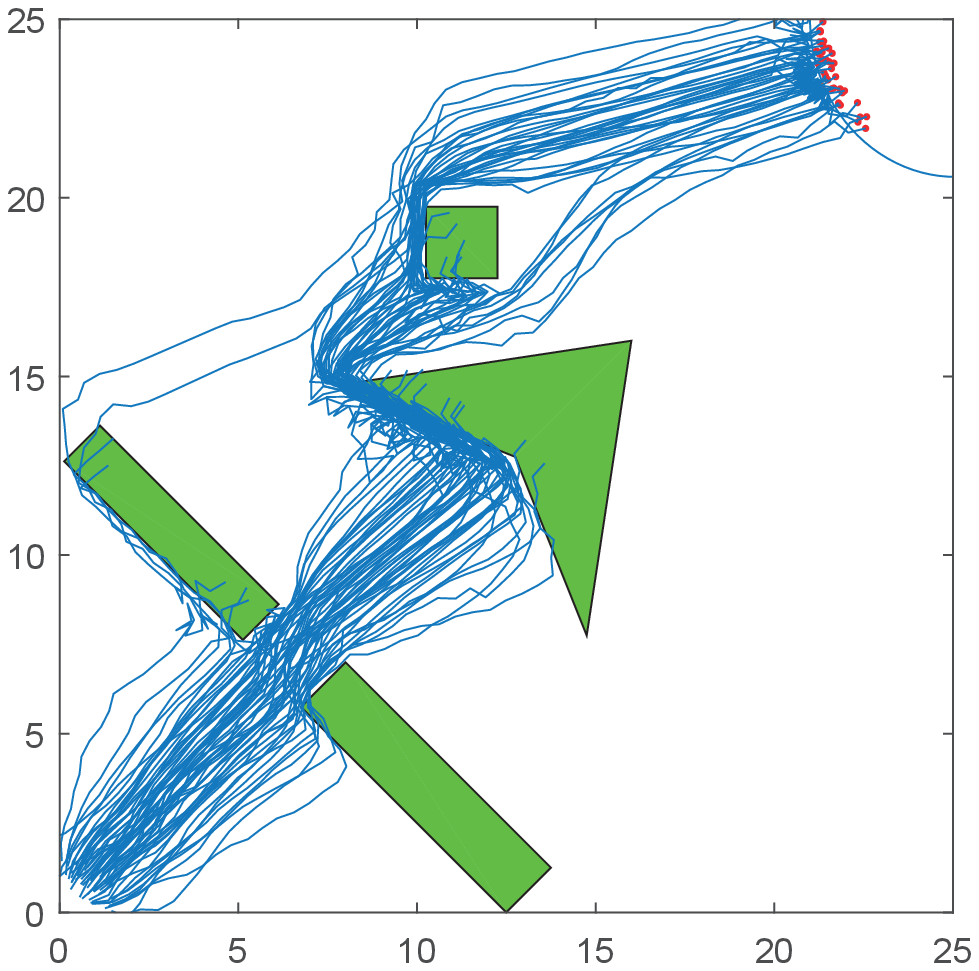}
	}
	\quad
	\subfigure[BRIAN (60 robots)]{
		\includegraphics[width=3.6cm]{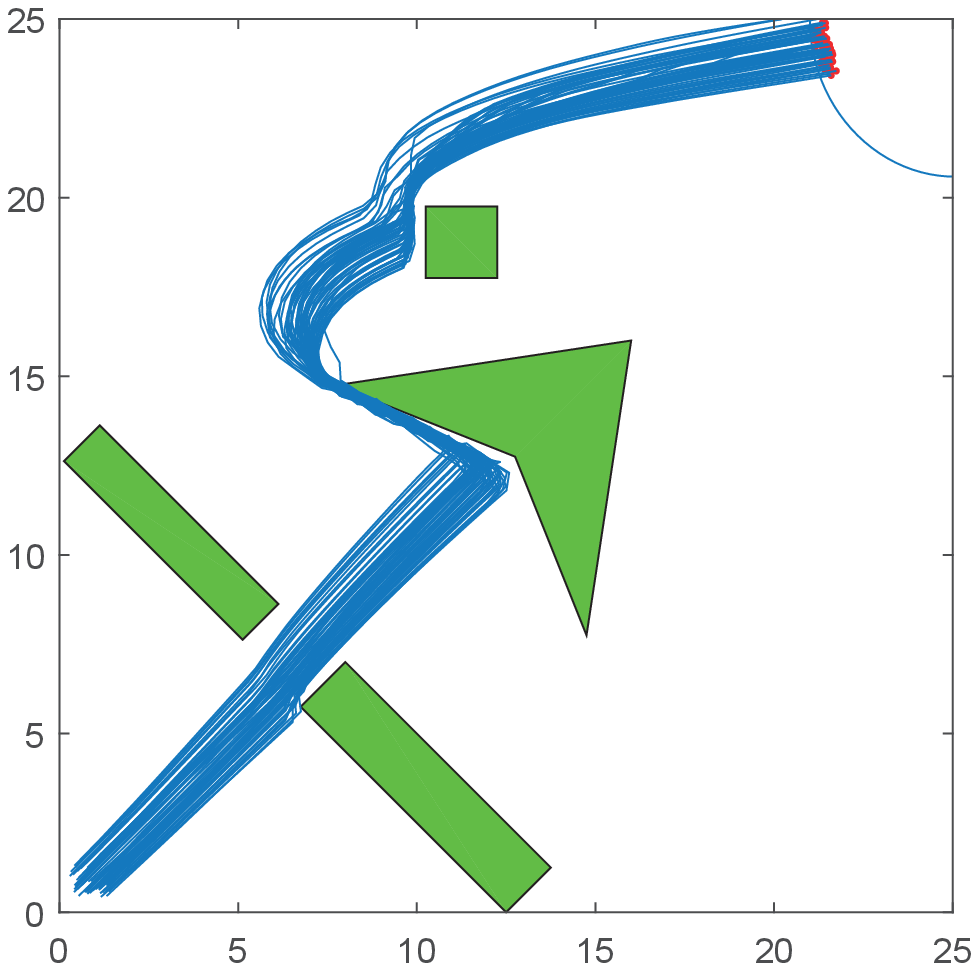}
	}
	\quad
	\subfigure[BREAM (100 robots)]{
		\includegraphics[width=3.6cm]{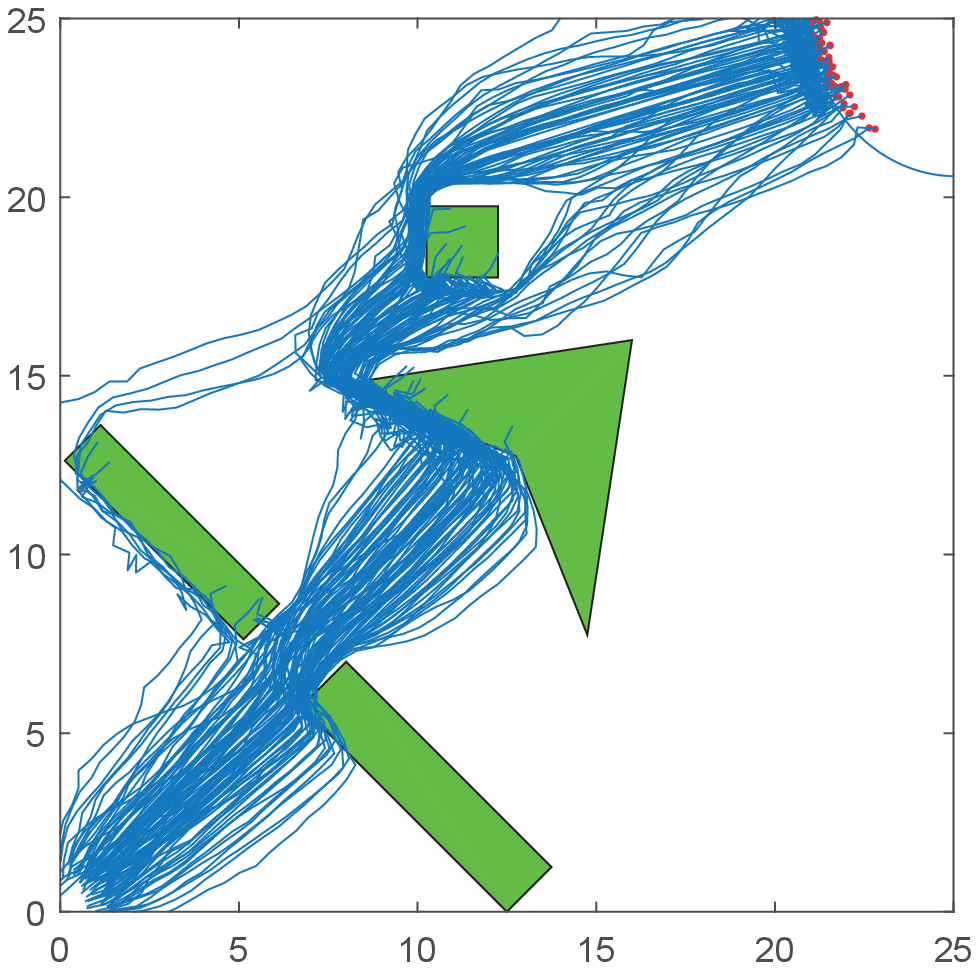}
	}
	\quad
	\subfigure[BRIAN (100 robots)]{
		\includegraphics[width=3.6cm]{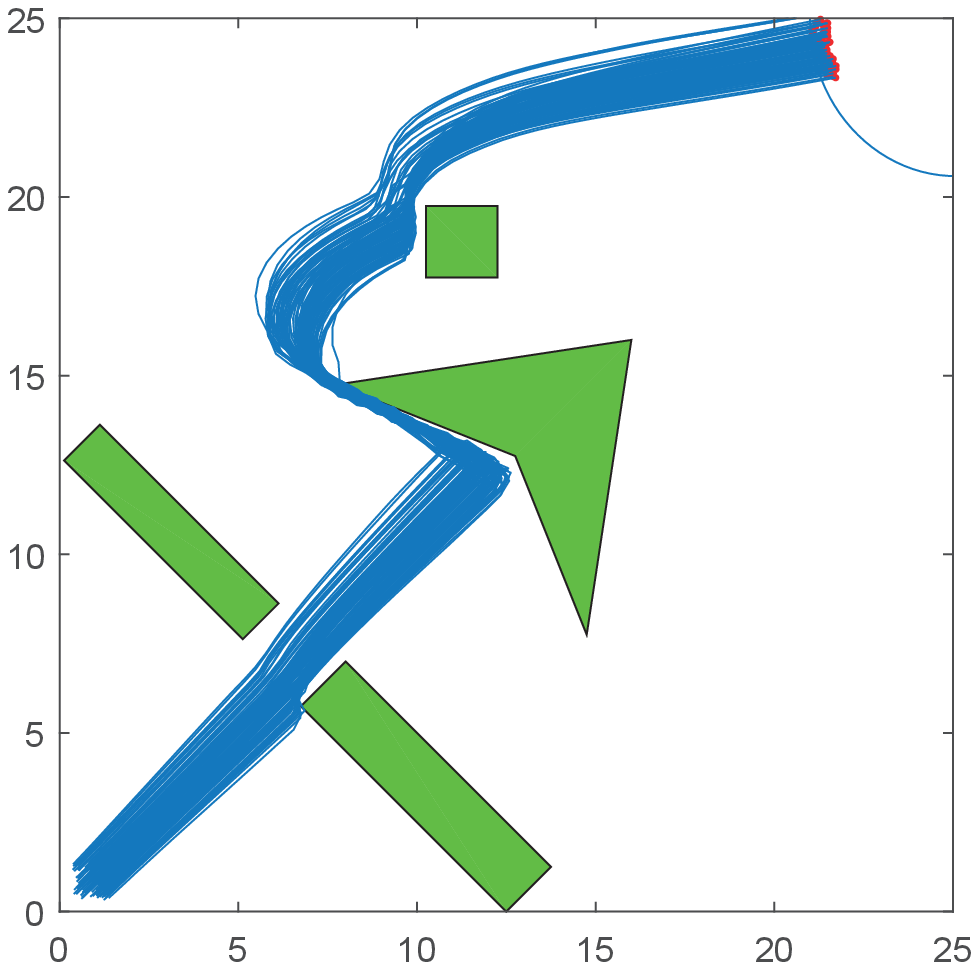}
	}
	\caption{ BREAM and BRIAN traces of the robotic swarm in autonomous navigation in test scenery ( We tested 3 groups of experiments using 20, 60 and 100 robotic agents for simulation.)}
\end{figure}

In order to clearly observe the impacts of different parameters of the formula for velcocity updating, we compare the performance of basic rule-based model (BREAM) and our optimized flocking model for robotic swarm in navigation (BRIAN) in scenary with three basic environmental elements including tunnel obstacle, non-convex obstacle and convex obstacle.

\begin{table*}[!htb]
	\centering
	\caption{Comparisons between BREAM \& BRIAN with 20, 60 and 100 robots}
	\label{table}
	
	\setlength{\tabcolsep}{1.8mm}
	\begin{tabular}{cccccccccccccc}
		\hline
		Evaluation& 
		\multicolumn{2}{c}{ Aggregation }& 
		\multicolumn{2}{c}{ Anisotropy }&
		\multicolumn{2}{c}{ Averagetime }&
		\multicolumn{2}{c}{ Uniformity }&
		\multicolumn{2}{c}{ Deathrate }&
		\multicolumn{2}{c}{ Fitness Function } \\
		\hline
		Algorithm& 
		BREAM&
		BRIAN&
		BREAM&
		BRIAN&
		BREAM&
		BRIAN&
		BREAM&
		BRIAN&
		BREAM&
		BRIAN&
		BREAM&
		BRIAN \\
		\hline
		Num-20& 
		0.8481&
		0.4666&
		38.6869&
		5.3094&
		130.0500&
		84.5000&
		0.2864&
		0.0076&
		0.3500&
		0.0000&
		6.1110&
		0.0079 \\
		\hline
		Num-60& 
		0.8639&
		0.4326&
		42.8201&
		4.6934&
		128.7667&
		83.9167&
		0.3195&
		0.0435&
		0.3333&
		0.0000&
		7.6107&
		0.0371 \\
		\hline
		Num-100& 
		1.1681&
		0.4557&
		50.1455&
		4.9927&
		115.8000&
		84.2500&
		0.2100&
		0.0330&
		2.7367&
		0.0000&
		92.8195&
		0.0317 \\
		\hline
	\end{tabular}
	\label{tab1}
\end{table*}

Fig. 3 shows directly that BRIAN performs better than BREAM in uniformity and stability. Fig. 3(a) and Fig. 3(b) represent the performance of these two models with 20 robotic agents, Fig. 3(c) and Fig. 3(d) with 60 robots, Fig. 3(e) and Fig. 3(f) with 100 robots. Specific performance indicators are shown in Table \uppercase\expandafter{\romannumeral1}. We record the values of each evaluation index of the two models in three situations of the number and scale of robots. Generally, all the indicators of BRIAN model perform better (the smaller, the better). Specifically, aggregation of BRIAN is 56\% lower than that of BREAM while the reduction of other indicators (anisotropy, averagetime, uniformity, deathrate, and fitness function) are 88.61\%, 32.55\%, 89.69\%, 100\%, and 99.92\%, respectively.

\begin{figure}[H]
	\centerline{\includegraphics[width=\columnwidth]{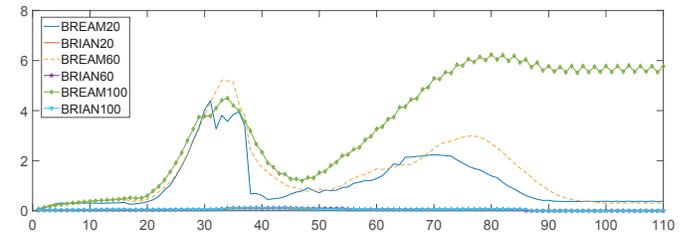}}
	\caption{ The uniformity of the robotic swarm with each experiment changes through time.}
	\label{fig5}
\end{figure}

Fig. 4 shows the change of uniformity in the whole time step. The total time step of each group of experiments is not the same, but it can be seen from the figure that the data of each group of BRIAN are stable between 0 and 1, which means that the stability and tightness of the cluster are very good during the whole cruise. When BREAM passes through obstacles, it can be seen that there will be large fluctuations near step 31 and step 71. Such fluctuations represent the situation of low cluster tightness and stability when cluster passes through narrow and non-convex obstacles, and the formation is not well maintained. At the same time, it can be seen that BRIAN has completed the whole task in about 84 seconds, while BREAM has completed the whole task.

\section{Conclusions and future work}

We presented in this paper an optimized flocking model for robotic swarm in autonomous navigation. This model is obtained through G-flocking algorithm proposed by us, which is extended from the classical genetic algorithm and rule-based flocking model in most relative researches. This is the first of its kind reported in the literatures; it comprehensively addresses the reliability, adaptivity and scalability of the robotic swarm during completing the navigation tasks.

The following issue will be addressed in our future work: First,  we will extend our experiment to the real-world systems such as unmanned aerial systems and unmanned ground systems. Second, we will take more uncertainties of scenaries into the model to verify the correctness of our model, such as adding the moving obstacle, the irregular barriers, and even fluid barriers.

\begin{IEEEbiography}[{\includegraphics[width=1in,height=1.25in,clip,keepaspectratio]{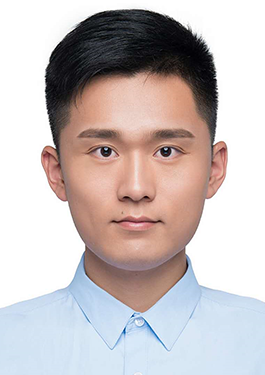}}]{Li Ma} is currently working toward his Ph.D. degree in the College of Systems Engineering, National University of Defense Technology. Contact him at mali10@nudt.edu.cn.
\end{IEEEbiography}
\begin{IEEEbiography}[{\includegraphics[width=1in,height=1.25in,clip,keepaspectratio]{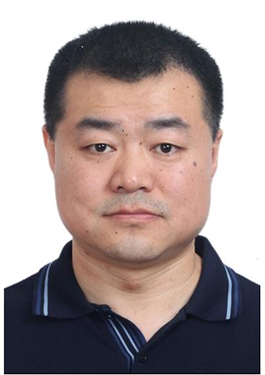}}]{Weidong Bao} is currently a professor in the College of Systems Engineering at National University of Defense Technology, Changsha, China. Contact him at wdbao@nudt.edu.cn.
\end{IEEEbiography}
\begin{IEEEbiography}[{\includegraphics[width=1in,height=1.25in,clip,keepaspectratio]{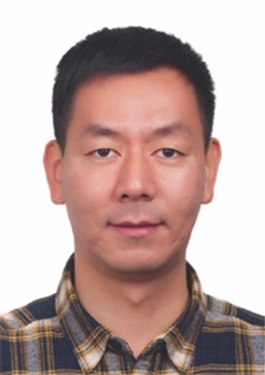}}]{Xiaomin Zhu} is currently an Associate Professor in the College of Systems Engineering at National University of Defense Technology, Changsha, China. Contact him at xmzhu@nudt.edu.cn.
\end{IEEEbiography}
\begin{IEEEbiography}[{\includegraphics[width=1in,height=1.25in,clip,keepaspectratio]{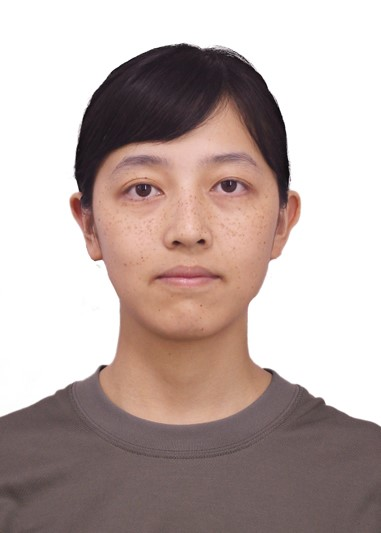}}]{Meng Wu} is currently pursuing the M.S. degree in the College of Systems Engineering, National University of Defense Technology, China. Contact her at wumeng15@nudt.edu.cn.
\end{IEEEbiography}
\begin{IEEEbiography}[{\includegraphics[width=1in,height=1.25in,clip,keepaspectratio]{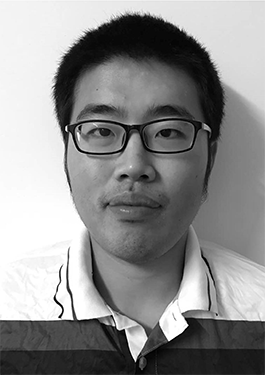}}]{Yuan Wang} is currently a Ph.D Candidate of College of Systems Engineering, National University of Defense Technology, Changsha, China. Contact him at wy1020395067@hotmail.com.
\end{IEEEbiography}
\begin{IEEEbiography}[{\includegraphics[width=1in,height=1.25in,clip,keepaspectratio]{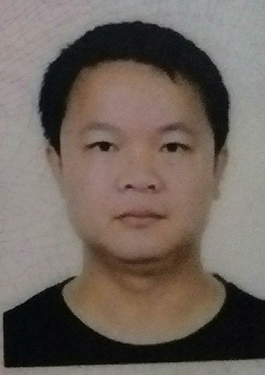}}]{Yunxiang Ling} is currently a professor in Officers college of PAP, Chengdu, China. Contact him at 2923821396@qq.com.
\end{IEEEbiography}
\begin{IEEEbiography}[{\includegraphics[width=1in,height=1.25in,clip,keepaspectratio]{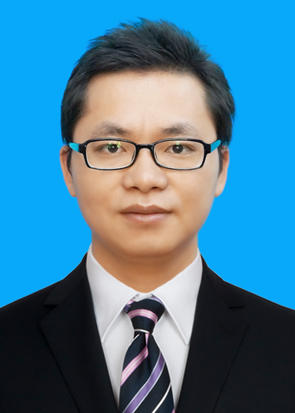}}]{Wen Zhou} is currently an Assistant Professor in the College of System Engineering at National University of Defense Technology, Changsha, China. Contact him at zhouwen@nudt.edu.cn.
\end{IEEEbiography}
\end{document}